\documentclass[twocolumn,showpacs,showkeys,amsmath,amssymb,aps,prl]{revtex4-1}
\usepackage[final]{graphicx} % to include figures
\usepackage{natbib}
\graphicspath{{./figs/}}
\usepackage{psfrag}
\usepackage[usenames,dvipsnames]{color}

\newcommand{\Rm}{\mbox{\it Rm}}

\newcommand{\bnabla}{\mbox{\boldmath $\nabla$}}
\renewcommand{\vec}[1]{\mbox{\boldmath $#1$}}

\newcommand{\scL}{\mathcal{L}}

\begin{document}

%\title{On the optimised magnetic dynamo}
\title{Optimisation of the magnetic dynamo}

\author{Ashley P. Willis}

\affiliation{
   School of Mathematics and Statistics,
   University of Sheffield, S3\,7RH, U.K.}

\date{\today}

\begin{abstract}
In stars and planets, magnetic fields are believed to originate from
the motion of electrically conducting fluids in their interior,
through a process known as the dynamo mechanism.
In this letter,
an optimisation procedure
%optimisation of the velocity field for magnetic energy growth
%enables us 
is used 
to simultaneously address two fundamental questions of 
dynamo theory: 
``Which velocity field leads to the most magnetic energy growth?'' and 
``How large does the velocity need to be relative to magnetic diffusion?''.
In general, this requires optimisation over 
%Here 
the full space of continuous solenoidal velocity fields
possible within the geometry.  Here the case of a periodic box is considered.
Measuring the strength of the flow with the 
%A frequently used measure of the strength of the flow is the 
root-mean-square amplitude,
an optimal velocity field is shown to exist, but without 
limitation on the strain rate, 
optimisation is prone to divergence.
%it is locally optimal only.
Measuring the flow in terms of its associated dissipation leads to
the identification of a single optimal at the critical
magnetic Reynolds number necessary for a dynamo.
This magnetic Reynolds number 
is found to be only 15\% higher than that necessary for transient
growth of the magnetic field.
%, implying that stationary velocity fields
%are sufficient at low magnetic Reynolds number.
%%
%This minimum $\Rm$ for long-term growth is only 15\% 
%above that required for transient growth.  Time dependent velocities,
%which can exploit transient growth, therefore do not appear to be 
%essential at such low $\Rm$.

\end{abstract}


\maketitle 

The continuous stretching and folding of magnetic field lines
by a velocity field is considered to be the main mechanism 
generating magnetic fields in stars, planets and the interstellar
media \cite{Moffatt78}.  
This magnetic dynamo mechanism must counter magnetic 
diffusion, which occurs on a time scale that can be estimated by 
$L^2/\lambda$, where $L$ is the length scale of the system
and $\lambda$ is the magnetic diffusivity.
Together with a scale for the velocity $U$, the relative growth versus 
diffusion can be estimated with the magnetic Reynolds number,
$\Rm=LU/\lambda$.
Without complete knowledge of the interior flows of astrophysical
bodies, theoretical studies have considered many parametrised 
velocity fields in several geometries, including the 
Ponomorenko \cite{Gilbert88},
Roberts \cite{Roberts70}, 
Arnold-Beltrami-Childress
(ABC) \cite{Arnold65,Childress70} 
and
Dudley and James flows \cite{Dudley89},
over the last 40 years.
Systematic searches continue to seek the best dynamo possible  
within the parameter space, e.g.\ \cite{Alexakis11}.

The small length scale of the laboratory compared to astrophysical
bodies implies rapid diffusion, and the mechanism is therefore
difficult to reproduce. Nevertheless, this is an exciting era 
where laboratory experiments
have begun to realise this process
\cite{Gailitis01,Steiglitz01,Monchaux07}.
%and new experiments are planned in the U.S. and Switzerland
%ANY CITATIONS?.
The experiments vary greatly in geometry, 
but in order to be successful, all seek to optimise
the flow conditions necessary to realise magnetic energy growth.  

In this letter, optimisation is shown to be possible
without need for the specification of a parametrised
set of acceptable flows.  This enables a lower bound on the magnetic 
Reynolds number to be identified for a dynamo.
%Optimising for both long- and short-term
%growth reveals that the dynamo mechanism need not rely on 
%transient growth effects to work well, and that a stationary
%velocity field is sufficient at low $\Rm$.  

Throughout this letter, the length scale is taken to be $L=L_x/(2\pi)$, 
so that the scaled box has length $2\pi$ in each direction, 
and the following notations are used:
\begin{equation}
   \vec{v}^2 = \vec{v}\cdot\vec{v}, \quad
   \langle a \rangle = \frac{1}{V}\int a \, \mathrm{d}V , \quad
%   ||a||_2 = \langle a^2 \rangle^{\frac{1}{2}},
   ||a||_n = \langle a^n \rangle^{1/n}.
\end{equation}
$V$ is the volume of the box,
so that $||\cdot||_2$ is equivalent to the root-mean-square value.
To begin with, the velocity scale is taken to be $U=||\vec{u}||_2$.

A variational optimisation method is used to find the 
velocity field 
$\vec{u}=\vec{u}(\vec{x})$ that maximises the growth 
in the magnetic field 
$\vec{B}=\vec{B}(\vec{x},t)$
after a period of time $T$.  
This method has recently proven useful in the study of the growth
of disturbances in shear flows \cite{Zuccher06,Pringle10}.
Consider the objective function
\begin{eqnarray}
\label{eq:L}
\scL & = & \langle \vec{B}_T^2 \rangle 
\,-\, \lambda_1 \, ( \langle \vec{u}^2 \rangle - 1 )
\,-\, \lambda_2 \, ( \langle \vec{B}_0^2 \rangle - 1 ) \nonumber \\
& & - \langle \Pi_1 \,\bnabla\cdot\vec{u} \rangle
- \langle \Pi_2 \,\bnabla\cdot\vec{B}_0 \rangle \\
& &  
- \int_0^T \langle \vec{\Gamma} \cdot 
  [ \partial_t \vec{B} - \bnabla\times(\vec{u}\times\vec{B}) 
   - \frac{1}{\Rm}\,\bnabla^2\vec{B} ] \,
  \rangle \mathrm{d}t \, ,\nonumber 
\end{eqnarray}
where 
$\vec{B}_0=\vec{B}(\vec{x},0)$ and
$\vec{B}_T=\vec{B}(\vec{x},T)$.
The first term on the right-hand side is to be maximised.
The remaining terms are constraints, including 
Lagrange multipliers $\lambda_i$, 
$\Pi_i=\Pi_i(\vec{x})$ and $\vec{\Gamma}=\vec{\Gamma}(\vec{x},t)$.
These terms are enforced to be zero.
As the induction equation preserves the solenoidal condition
on $\vec{B}$, it is specified for $\vec{B}_0$ only.
After applying variational derivatives it may be written that
\begin{eqnarray}
\delta\scL & = & 
\langle \frac{\delta\scL}{\vec{\delta u}}\cdot\vec{\delta u}\rangle
+ \langle\frac{\delta\scL}{\vec{\delta B}_0}\cdot\vec{\delta B}_0\rangle
+ \langle\frac{\delta\scL}{\vec{\delta B}_T}\cdot\vec{\delta B}_T\rangle
\nonumber \\
& & 
- \int_0^T \langle \vec{\delta \Gamma} \cdot [ \mbox{ind.} ] \,
  \rangle \mathrm{d}t
- \int_0^T \langle \vec{\delta B} \cdot [ \mbox{adj.} ] \,
  \rangle \mathrm{d}t \, ,
\label{eq:Lvar}
\end{eqnarray}
where
\begin{eqnarray}
\label{eq:dLu}
\frac{\delta\scL}{\vec{\delta u}} ~~ & = & 
\int_0^T \vec{B}\times(\bnabla\times\vec{\Gamma})\,\mathrm{d}t 
- 2 \, \lambda_1\,\vec{u} + \bnabla\Pi_1 , 
\\
\label{eq:dLB0}
\frac{\delta\scL}{\vec{\delta B}_0} & = &
\vec{\Gamma}_0 - 2\,\lambda_2\,\vec{B}_0 + \bnabla \Pi_2 , 
\\
\label{eq:dLBT}
\frac{\delta\scL}{\vec{\delta B}_T} & = &
  2\,\vec{B}_T - \vec{\Gamma}_T ,
\end{eqnarray}
``ind.'' represents the induction equation, 
as it appears in (\ref{eq:L}), and
``adj.'' is set to zero giving the adjoint equation
\begin{equation}
   -\partial_t\vec{\Gamma} = (\bnabla\times\vec{\Gamma})\times\vec{u} 
   + \frac{1}{Rm}\bnabla^2 \vec{\Gamma} \, .
\end{equation}
In deriving these expressions it is necessary to lift
derivatives off the variations, for example
\begin{eqnarray}
\langle \Pi \, \bnabla \cdot\vec{\delta v} \rangle & = &
   \langle \Pi \, \partial_i \,\delta v_i \rangle = 
   \langle \partial_i \Pi\, \delta v_i \rangle 
   - \langle \delta v_i \, \partial_i \Pi \rangle \nonumber\\
& = & \frac{1}{V} \int \Pi\,\vec{\delta v}\cdot \vec{\mathrm{d}S}
   - \langle \vec{\delta v}\cdot\bnabla\Pi \rangle ,
\end{eqnarray}
where the product rule and Gauss' Theorem have been used.
For the first of the final two terms, 
the integral over the closed surface vanishes for case of the 
periodic box.  For the second, it is quite beautiful that the 
Lagrange multipliers themselves provide projection functions 
--- these will be used to ensure that $\vec{u}$ and $\vec{B}_0$ 
are solenoidal.

For a given $\vec{u}$ and $\vec{B}_0$, both solenoidal and 
normalised, timestepping the induction equation to give $\vec{B}_T$
ensures that the penultimate term on the right-hand side of 
(\ref{eq:Lvar}) is zero.  Then $\partial\scL/\vec{\delta B}_T$ 
in (\ref{eq:Lvar}) and (\ref{eq:dLBT}) is set to zero with
the compatibility condition
$\vec{\Gamma}_T=2\,\vec{B}_T$.  Given $\vec{\Gamma}_T$, 
timestepping the adjoint backwards sets the last term in
(\ref{eq:Lvar}) to zero and provides $\vec{\Gamma}_0$.
Now all quantities are known to calculate ascent directions for
$\scL$ given by (\ref{eq:dLu}) and (\ref{eq:dLB0}).
New fields that lead to an increased $\scL$ are given by 
$\vec{u}:=\vec{u} + \epsilon \, (\delta\scL/\vec{\delta u})$
and
$\vec{B}_0:=\vec{B}_0 + \epsilon \, (\delta\scL/\vec{\delta B}_0)$,
where $\epsilon$ is a small scalar value.
The new fields are projected onto the space of solenoidal 
functions by considering the divergence of the update,
which defines the projection functions $\Pi_i$.
After this, the $\lambda_i$ are then chosen such
that the new fields have unit norm, and all constraint
terms in (\ref{eq:L}) are then zero.  
Further details regarding a similar implementation of the 
method for pipe flow can be found in \cite{Pringle12}.

The value of $\epsilon$ is adjusted according 
to whether or not consecutive updates appear to be 
pointing in a similar
direction.  Note also that
to evaluate (\ref{eq:dLu}), $\vec{B}$ needs to be known for all 
intermediate times
during the backwards integration of $\vec{\Gamma}$.  This could require
significant computer memory.  Instead $\vec{B}$ may be saved at 
`checkpoints' and re-integrated forwards when needed, involving
only 50\% extra work overall.
Spatial discretisation used in the time-stepping code 
is via a triple-Fourier expansion.
%% FORMULATION MOVED TO END
Nonlinear terms are evaluated pseudospectrally on a grid with 
at least 36 points in each direction.  This resolution was found to
be more than sufficient for the majority of calculations, where $\Rm$ 
is very low.  
%It has been verified that the code accurately reproduces 
%growth rates for several configurations of 
%the ABC flow \cite{sdfdf}.

\begin{figure}
   \centering
   \psfrag{ET}{\small $\langle\vec{B}_T^2\rangle$}
   \psfrag{s}{\small $\sigma$}
   \psfrag{d}{\small $d$}
   \psfrag{iteration}{\small iteration}
   \psfrag{error}{\small err.}
   \includegraphics[width=0.80\linewidth]{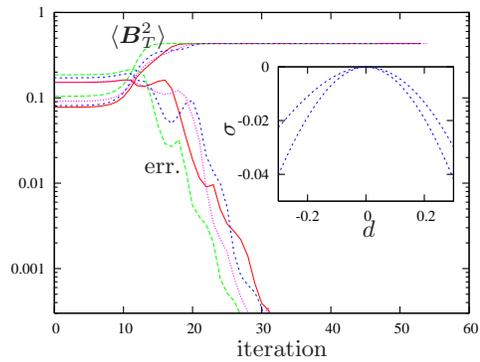}
   \caption{\label{fig:cgce}
      Several initial conditions converge to the same optimal;
      $\Rm=1$, $T=1$.  
      {\em Inset}: 
      Sensitivity of the optimal $\vec{u}_{op}$ at $\Rm_c=1.737$
      to perturbations
      $\vec{u}_p$ measured by the growth rate, 
      $\sigma$, in units $||\vec{u}||_2/L$.  Here 
      $\vec{u}=\alpha(\vec{u}_{op}+d\,\vec{u}_p)$,
      where $\vec{u}_p$ is randomly chosen, 
      $||\vec{u}_{op}||_2=||\vec{u}_p||_2=1$, and 
      $\alpha$ is such that $||\vec{u}||_2=1$.
   }
\end{figure}
Figure \ref{fig:cgce}
shows the result of optimisations starting from several 
random initial $\vec{u}$ and $\vec{B}_0$ at $\Rm=1$.  
All converge to the same optimal, and 
the error, measured by 
$(\langle(\delta\scL/\vec{\delta u})^2\rangle +
\langle(\delta\scL/\vec{\delta B}_0)^2\rangle)^{1/2}$
drops by 5 orders of magnitude during the calculation.

%\begin{figure}
%   \centering
%   \psfrag{E}{\small $\langle\vec{B}^2\rangle$}
%   \psfrag{t}{\small $t$}
%   \psfrag{it}{\small iter.}
%   \includegraphics[width=0.9\linewidth]{figs/3808Bt}
%   \caption{\label{fig:Bt}
%      Magnetic energy for converged optima at $\Rm=1.5$, $1.737$ 
%      and $2.0$.  By time $T$ the signal is dominated 
%      by the leading eigenmode.
%      {\em Inset}:
%      Divergence for $\Rm=2.2$; quantities from bottom to top 
%      are $\sigma$,  $||S||_2=||\vec{\omega}||_2$,
%      $||\vec{\omega}||_\infty$ and $||S||_\infty$,
%   }
%\end{figure}
%Figure \ref{fig:Bt} shows the dependence of 
%$\langle \vec{B}^2 \rangle$ with time.
Increasing $\Rm$, the optimal velocity field
is found to change little until the growth rate, $\sigma$, 
of the magnetic field is zero at
$\Rm_c=1.737$ for $\vec{u}=\vec{u}_{op}$.
%The growth rate is calculated at $T=7$,
At low $\Rm$, very modest transient growth is observed.
For $\Rm=\Rm_c$, transient growth of only 
2\% occurs, all within the first 2 time units.
By the time $T=7$ is reached, 
the field $\vec{B}_T$ is dominated by the leading eigenmode
and the energy is steady; $\sigma$, calculated at
the end time $T$, is zero in this case.
In this non-dimensionalisation the diffusion time 
$L^2/\lambda$ is equal to $t=\Rm$ in units $L/||\vec{u}||_2$.
The sensitivity to perturbations of the optimal is shown in the 
inset to Fig.\ \ref{fig:cgce}.
The optimal is apparently quite robust at low $\Rm$.
A decrease in the growth rate for all perturbations
also serves as a good test that the calculated velocity field
is indeed optimal.

\begin{figure}
  \centering
  \includegraphics[width=0.85\linewidth]{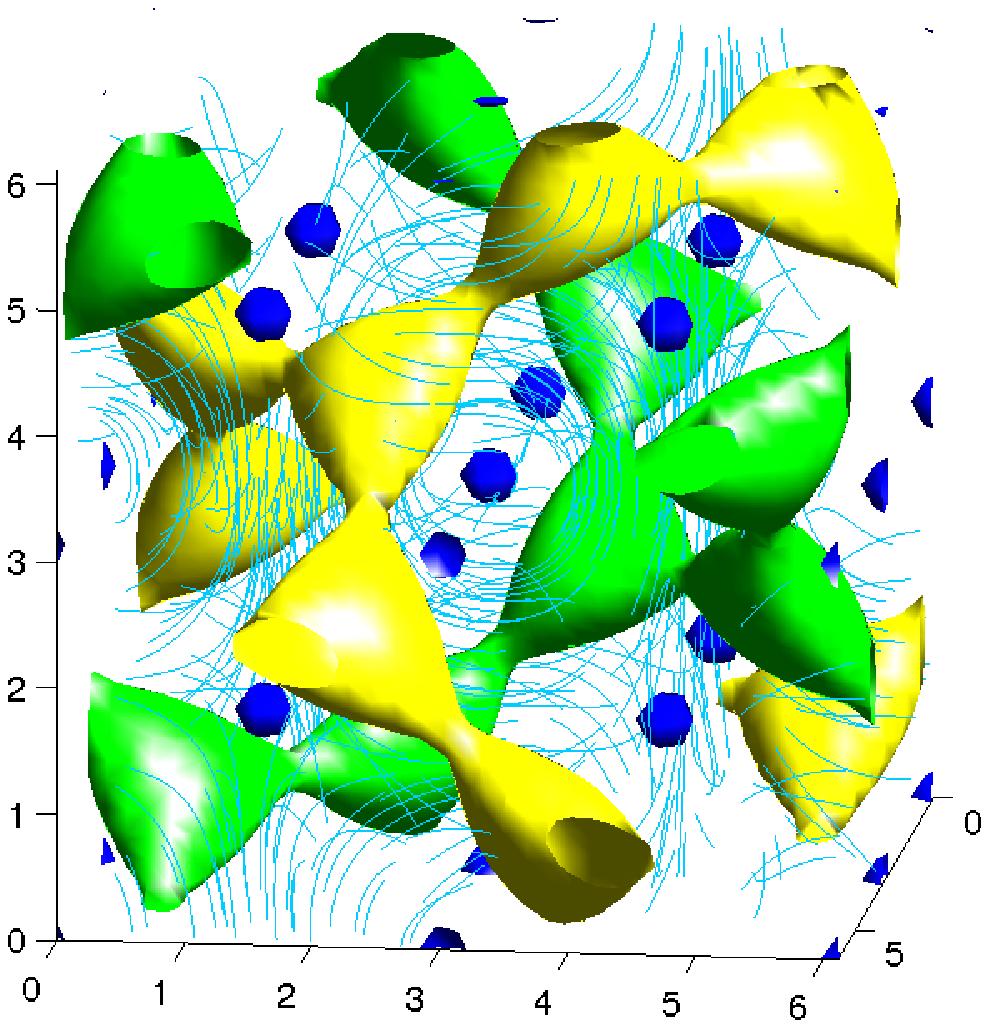}\\[-10pt]
  ({\it a})\\[-2pt]
  \includegraphics[width=0.85\linewidth]{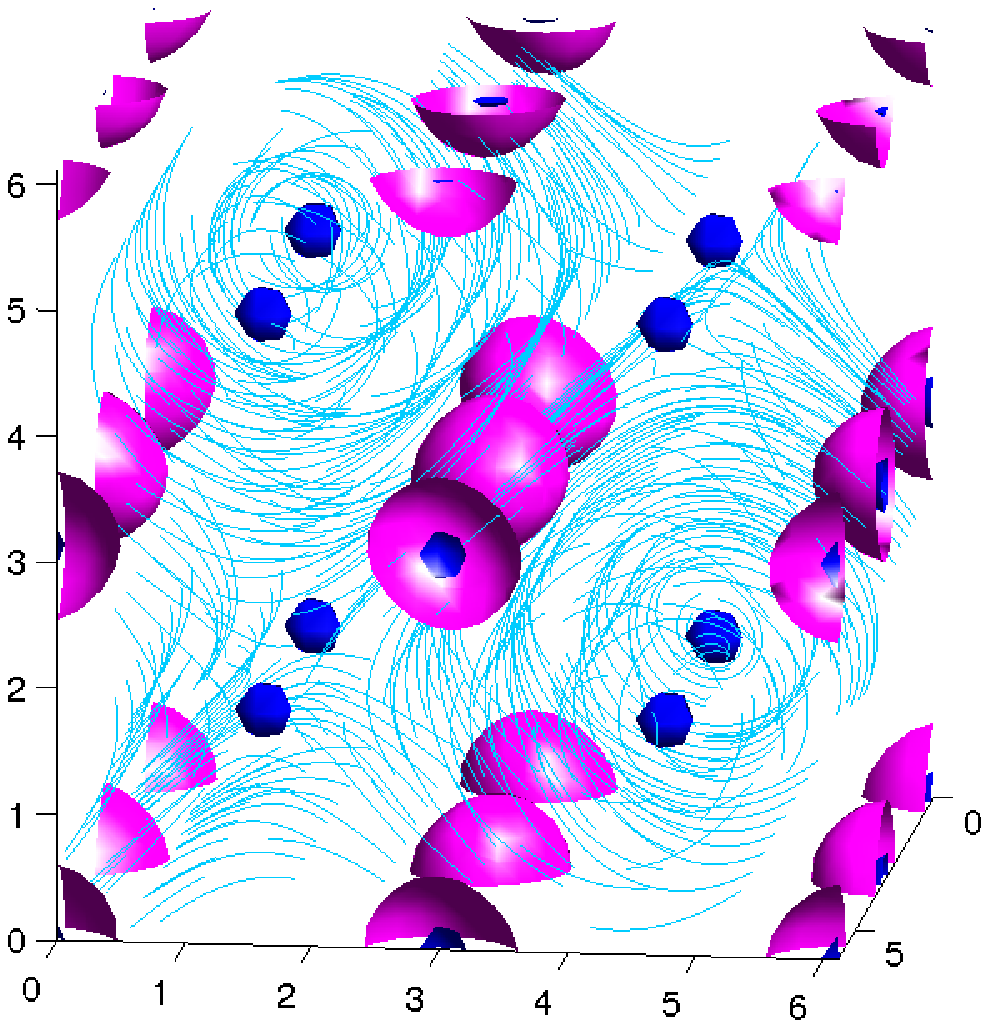}\\[-10pt]
  ({\it b})
  \caption{\label{fig:bflow} %(color online) 
     Optimal at $\Rm_c=1.737$\,.
     ({\it a}) Isosurfaces of positive and negative helicity 
     $\vec{u}\cdot{\vec{\omega}}$ (yellow, green)
     and stagnation points (blue).
     ({\it b}) Isosurfaces of $\vec{B}_T^2$ (pink).
     Blue lines in ({\it a}) and ({\it b}) are respectively
     streamlines and magnetic field lines.
  } 
\end{figure}

Helicity is plotted 
in Fig.\ \ref{fig:bflow}
to give a sense of the geometry and symmetry
of the flow. Its mean helicity
$\langle\vec{u}\cdot\vec{\omega}\rangle$ 
%%/ (||\vec{u}||_2\,||\vec{\omega}||_2)$,
is zero, but the maximum $\vec{u}\cdot\vec{\omega}$
over the domain is high at $3.88$.  The flow has a 
measured length scale \cite{Alexakis11} 
$k_u=||\vec{\omega}||_2/||\vec{u}||_2=1.478$
and is therefore large-scale.
The magnetic field is centred on a 
subset of the stagnation points in the flow;
see \cite{Archontis03} for plots where the dynamo
mechanism appears to be similar.  
As $\vec{u}_{op}$
is dominated by the largest length scale, a simple 
approximation is possible, given by
\begin{equation}
   \vec{u}_{op} \approx \vec{u}_a = \frac{2}{\sqrt{3}}\,%(2/\sqrt{3})\, 
   \left(\sin y\cos z, \, \sin z\cos x, \, \sin x\cos y\right) 
   \, .
\end{equation}
The maximum helicity 
%$\vec{u}\cdot\vec{\omega}$ for the approximation
for $\vec{u}_a$, at 2, is less than that for $\vec{u}_{op}$,
and the relative difference between the fields is
$||\vec{u}_{op}-\vec{u}_a||=0.16$.  Despite this,
$\Rm_c=1.761$ is only slightly elevated for
the approximation.

Although the optimal can be traced a little
beyond $\Rm=2$, it becomes clear that its 
basin of optimality (with respect to $\scL$)
quickly becomes vanishingly small,
as not all optimisations converge.
When the optimisation converges, the velocity and
magnetic field are well resolved, and the energy spectrum
of the Fourier coefficients
falls by 16 orders.  
But for only slightly larger $\Rm$, 
other velocity fields that
include highly localised regions of strain are
also picked up by the optimisation, and
large strain is known to lead to large growth
\cite{Galanti92}.  
%Improved resolution does not remedy this problem, it only 
%permits even finer structures to form.
While the scaled velocity satisfies $||\vec{u}||_2=1$,
the strain rate $S=(2\,S_{ij}\,S_{ij})^{1/2}$,
where
$S_{ij}=\frac{1}{2}(\partial_i\,u_j+\partial_j\,u_i)$,
is unlimited.  Hence the growth is unlimited and
the optimisation fails.
%The inset to Figure \ref{fig:Bt} shows an example where the
%method deviates from the local optimal at $\Rm=2.2$.  

A more appropriate measure of the flow is therefore 
necessary, here taken to be the velocity scale
 $U=L\,||S||_2$.  Whilst $S$ and the vorticity
$|\vec{\omega}|$ are not equal locally, it can be 
shown that $||S||_2=||\vec{\omega}||_2$.  The latter is
somewhat more convenient to work with in the variational method.
The scaled velocity field then satisfies $||\vec{\omega}||_2=1$,
so that optimisation can be considered to be over the
space of fields with equal viscous dissipation or input power
driving the flow, neglecting any feedback on the flow from 
the magnetic field.
This scaling leads to the magnetic Reynolds number
\begin{equation}
   \Rm_\omega = \frac{L^2\,||\vec{\omega}||_2}{\lambda} .
\end{equation}
%where $S=(2\,S_{ij}\,S_{ij})^{\frac{1}{2}}$ and
%$S_{ij}=\frac{1}{2}(\partial_i\,u_j+\partial_j\,u_i)$.
In a similar context, 
Backus \cite{Backus58}
derived a lower limit for 
dynamos in a sphere in terms of the analogous magnetic Reynolds number 
except involving $||S||_\infty$ rather than $||S||_2$.
Although not used further here, higher norms of $S$ and $\vec{\omega}$ are 
observed to behave similarly.
%; see the inset to Figure \ref{fig:Bt}.

The optimisation is only slightly altered, where now
\begin{equation}
   \scL = \langle \vec{B}_T^2 \rangle 
   \, -\,  \lambda_1 (\langle \vec{\omega}^2 \rangle - 1) \,-\, \dots
\end{equation}
and the new update is given by
\begin{equation}
\frac{\delta\scL}{\vec{\delta u}} ~ = 
\int_0^T \vec{B}\times(\bnabla\times\vec{\Gamma})\,\mathrm{d}t 
%+ 2\lambda_1\, \bnabla^2\vec{u} + \bnabla\Pi_1 . \\
- 2\lambda_1\, \bnabla\times\vec{\omega} + \bnabla\Pi_1 . \\
\end{equation}
Similar to before, the scalar $\lambda_1$ is chosen such that 
for the new $\vec{u}$ one has $\langle\vec{\omega}^2\rangle=1$.
The extra complexity of the update leads to a linear
approximation that is valid over a shorter range, and 
the number of iterations necessary for convergence is 
typically order 1000, compared with order 100 before.

\begin{figure}
   \centering
   \psfrag{E}{\small $\langle\vec{B}^2\rangle$}
   \psfrag{t}{\small $t$}
   \psfrag{s}{\small $\sigma$}
   \psfrag{d}{\small $d$}
   \psfrag{RmS}{\small $\Rm_\omega$}
   \includegraphics[width=0.80\linewidth]{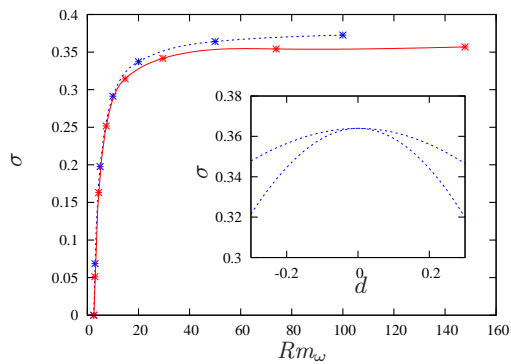}
   \caption{\label{fig:GrowthRe}
      Growth rates, $\sigma$ in units $||\vec{\omega}||_2$, for
      the optimal at $\Rm_c=1.737$ (red, solid), 
      where the velocity is held fixed.  This is bounded above by 
      the optimisation at each $\Rm_\omega$ (blue, dashed);
      $\Rm_{\omega c}=2.48$.        
      {\em Inset}:
      Growth rates for perturbations to the optimal at 
      $\Rm_\omega=50$ with $||\vec{\omega}||=1$; disturbances
      defined similarly in Fig.\ \ref{fig:cgce}.
%      Optimal initial growth at $Rm_{Sg}=2.12$.
%      The local optimal for which growth is initially zero defines 
%      $Rm_{Sg}$
%      (red, solid)
%      but decays rapidly at later times.  The other optimal 
%      monotonically decays (blue, dashed) but is 
%      connected to the optimal for long-term growth.
   }
\end{figure}

Upon optimisation with the vorticity scaling,
zero growth rate is found at $\Rm_{\omega c}=2.48$. 
The optimal velocity field is structurally almost identical
to that found previously at $\Rm_c=1.737$, but for the 
new velocity field, its corresponding $\Rm=1.75$, now an observed quantity, is slightly higher.
Starting from 40 initial conditions at $\Rm_{\omega c}$, all converged
to the same optimal.
%in any optimisation where
%$||\vec{\omega}||_2=1$, even starting 
%with diverged states from the $||\vec{u}||_2=1$ optimisations.
%%
Figure \ref{fig:GrowthRe} compares growth rates for the optimal found at 
$\Rm_c=1.737$, where $\vec{u}_{op}$ is fixed and $\Rm_\omega$ 
is varied,
with the optimal that could now be 
tracked up to $\Rm_\omega=100$.  For this range of
$\Rm_\omega$ the fixed velocity field is competitive
with the optimised state, which changes relatively little.
From $\Rm_\omega=2.48$ to $100$ it remains {\em an} optimal with 
only a relative change of 8\%.  
Although the inset shows that the optimal at $\Rm_\omega=50$ remains 
fairly robust to perturbations, at these larger 
magnetic Reynolds numbers, other
optimals are likely to exist.

Beyond $\Rm_\omega=100$, convergence was found to be possible 
using a higher norm $||\vec{\omega}||_4$, but this incurs further
iterations for convergence.  Optimisation at higher $\Rm$ 
becomes computationally expensive --- in addition to increased
spatial resolution and more iterations, 
larger target times $T$ are necessary to pass longer transients.

\begin{figure}
   \centering
   \psfrag{E}{\small $\langle\vec{B}^2\rangle$}
   \psfrag{t}{\small $t$}
   \includegraphics[width=0.62\linewidth]{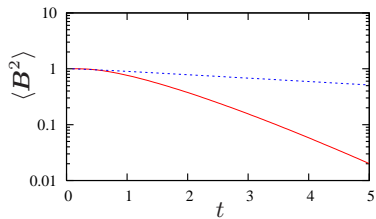}
   \caption{\label{fig:initGr}
      Development of energy following optimised initial growth;
      $Rm_{\omega g}=2.12$.
      The local optimal for which growth is initially zero defines 
      $Rm_{\omega g}$ (red, solid), but rapid decay immediately 
      follows the initial `burst'.  A second optimal (blue, dashed)
      decays monotonically but more slowly.
      This second optimal is connected to that for a steady 
      field at the slightly higher $\Rm_{\omega c}=2.48$.
   }
\end{figure}

Time dependence of the velocity field is an important factor that
could affect magnetic energy growth.  In order for energy growth to be
enhanced, the changing velocity field must exploit transient energy
growth to beat the mean of the energy growths associated with
each velocity field considered separately.  At low $\Rm$, however,
this affect has barely been observed.
Setting $T$ to a small value permits calculation of velocity and 
magnetic fields that lead to the largest initial magnetic energy
growth.  
Starting the optimisation from $40$ initial conditions and
taking $T=0.05$, 
two optima were identified, shown in
Fig.\ \ref{fig:initGr}.
The lowest $\Rm_\omega$ for which the maximum initial
growth was zero set the energy stability bound at 
$\Rm_{\omega g}=2.12$.  Between $\Rm_{\omega g}$ and 
$\Rm_{\omega c}$ it is possible to find brief growth
of the magnetic field, but ultimately it decays.
The small difference between
$\Rm_{\omega g}$ and $\Rm_{\omega c}$ leaves little room
for reduction of $\Rm_{\omega c}$ gained by 
time-dependence of velocity fields.
Transient growth will, however, be very
important at large $\Rm$.

In summary, 
a minimum magnetic Reynolds number is found for
a kinematic dynamo at $\Rm_{\omega c}=2.48$, by
optimisation over the space of velocity fields 
with equal $||\vec{\omega}||_2$ or associated viscous dissipation.
Starting from 40 random initial conditions at this $\Rm_\omega$, 
all converged to the same optimal.  
%For such a low $\Rm_\omega$ 
%the hypersurface for $\scL$ is expected to be relatively smooth,
%strongly suggesting that the optimal is unique.
This velocity field is very close to an
optimal at $\Rm_c=1.737$
in the space of fields with $||\vec{u}||_2=1$
\footnote{
For comparison, the space of ABC flows is possibly a 
restrictive space of flows, but at the same time it is 
sufficiently large to make the parameter search a 
substantial feat.  
Within this class it has nevertheless been 
possible to identify the Roberts flow as being optimal
\cite{Alexakis11}.  For this geometry 
($z$-wavenumber $=1$) the critical magnetic Reynolds number
for Roberts flow is $\Rm_c=8.79$.
For the traditional ABC flow ($A=B=C=1$), $\Rm_c\approx 15$.
%Optimisation is desirable, both to reduce the effort of 
%the search and to reduce the estimated $\Rm_c$.
  }.
In the latter space, other fields with high strain-rate are 
also possible.
The optimal velocity field appears to be a fast dynamo,
with a growth rate 
$\sigma=0.358\,||\vec{\omega}||_2 = 0.529\,||\vec{u}||_2/L$,
but optimisation at larger $\Rm$ is likely to lead to the 
identification of more efficient dynamos.
A lower bound for instantaneous magnetic energy growth is found 
at $\Rm_{\omega g}=2.12$, not far below $\Rm_{\omega c}=2.48$.

As shown here, the variational method can be used to
identify velocity fields that maximise magnetic 
energy growth, and to determine a lower bound on the 
critical Reynolds number for a dynamo.
It is a rare occasion that we can put a numerical 
figure to such an important parameter.
In geometries closer to
experiments, tricky boundary conditions on the magnetic field
for spheres and cylinders pose technical challenges, but 
it will be worthwhile overcoming them.  It will be interesting to
see if velocity fields realisable in experiment are close, 
or can be made closer, to optimal.  That the growth rate is 
observed to change little for perturbations about the optimal, 
is promising for the dynamo's robustness. 
It will also be of interest to further
assess the mechanism that optimises growth, and  
future
optimisations at higher $\Rm$ may identify alternative 
optimals and dynamo mechanisms to that observed here.

\begin{acknowledgments} 
  The author thanks 
  Chris Pringle for an introduction to the variational method,
  Eun-Jin Kim for helpful discussions,
  and the referees for insightful suggestions.
\end{acknowledgments} 

\bibliography{mhd.bib}

\end{document}